\documentclass[
aps,%
12pt,%
final,%
notitlepage,%
oneside,%
onecolumn,%
nobibnotes,%
nofootinbib,
superscriptaddress,%
noshowpacs,%
centertags]%
{revtex4}
\begin{document}
\selectlanguage{english}

%
%






\title{Double Beta Decay: Historical Review of 75 Years of Research
}



\author{\firstname{A.~S.}~\surname{Barabash}}
\email[E-mail:~]{barabash@itep.ru}
\affiliation{Institute of Theoretical and Experimental Physics, B.\
Cheremushkinskaya 25, 117259 Moscow, Russia}

\begin{abstract}
Main achievements during 75 years of research on double beta
decay have been reviewed. The existing experimental data have
been presented and the capabilities of the next-generation
detectors have been demonstrated.
\end{abstract}

\maketitle

\section{INTRODUCTION}

The term "double beta decay" appeared 75 years ago. In view of
this anniversary, I want to recall the main stages of the
investigation of double beta (2$\beta$) decay and to review the
main achievements in chronological order. In the course of time,
many details (including important and interesting) are often
forgotten, accents change, contributions of individual
researchers are revised, etc. For this reason, young scientists
beginning investigations of 2$\beta$ decay do not necessarily
know earlier achievements, motivation, and persons responsible
for certain achievements. I began to study 2$\beta $ decay in
1982 and is one of the "senior" (but not oldest!) researchers
of 2$\beta$ decay continuing actively work in this field.
Furthermore, I was happy to meet and work with almost all leading
specialists in the field of 2$\beta$ decay. These are the
reasons for this review.

The review is organized as follows. Section~2 is devoted to the
appearance of the problem of 2$\beta$ decay, Section~3 presents
information on first experiments on the search for this process,
the history of achievements in the investigation of 2$\beta$
decay since 1960 is presented in Sections~4--6 in the
chronological order. Each section ends with a brief list of the
main achievements during the corresponding period and the list of
collaborations and individual scientists significantly
contributing to the advance in this field. Section~7 briefly
presents the current status and prospects of the investigation of
2$\beta$ decay in the near future. The tables 4-6 at the end of the
review summarizes the main (in my opinion) stages in the history
of the investigation of 2$\beta$ decay. The contribution of
Russian and Soviet scientists to the investigation of 2$\beta$
decay is discussed in the Conclusion.

When writing this review, I tried to be maximally unbiased and
give weighed estimates of obtained results (taking into account
the opinions of leading experts in the field of 2$\beta$ decay).
Nevertheless, it is not clearly free of author's opinion and
author's estimate of the importance of various experiments and
contributions of individual scientists.

\section{"Birth" OF 2$\beta $ DECAY}

The problem of 2$\beta $ decay appears immediately after Pauli's
hypothesis on the existence of neutrino in 1930 and the
development of the theory of $\beta$ decay in 1933. This theory
was developed by Italian physicist Fermi by analogy with quantum
electrodynamics, but he introduced a new type of interactions,
weak interaction, to describe $\beta$ decay. This theory
provided good description of the $\beta ^{ - }$- and $\beta^{ +}$
decays of nuclei:
\begin{gather}
\label{eq1}
(A, Z) \to (A, Z + 1) + e^{-}+\bar {\nu },
\end{gather}
\begin{gather}
\label{eq2}
(A, Z) \to (A, Z - 1) + e^{ + }+\nu ,
\end{gather}
where $A$~is the atomic number, $Z$~is the charge number, $e^{ -
}$~is the electron, $e^{ + }$~is the positron, $\nu $~is the
neutrino, and $\bar {\nu }$~is the antineutrino. In 1935,
Goeppert-Mayer [1] for the first time pointed to the possibility
of the process of two-neutrino double beta (2$\beta$(2$\nu $))
decay , i.e., a process of the transformation of the ($A$, $Z$)
nucleus to the ($A$, $Z + 2$) nucleus with the emission of two
electrons and two (anti)neutrinos:
\begin{gather}
\label{eq3}
(A, Z) \to (A, Z + 2) + 2e^{-} + 2\bar {\nu }
\end{gather}
It is interesting that Wigner proposed to consider such a
possibility (as mentioned by Goeppert-Mayer [1]). In 1937,
Majorana [2] theoretically showed that the conclusions of the
theory of $\beta $ decay remain unchanged under the assumption of
the existence of only one type of the neutrino having no
antiparticle (i.e., $\nu \equiv \bar {\nu }$). Thus, the notion
of the Majorana neutrino (in contrast to the Dirac neutrino)
appeared, but the term Majorana neutrino appeared later. At the
same year, Racah [3] was the first who pointed out that such a
neutrino emitted, e.g., in process (1) can induce the reaction
\begin{gather}
\label{eq4}
\nu + (A, Z) \to (A, Z + 1) + e^{-}.
\end{gather}
This reaction cannot be induced by a Dirac neutrino (i.e.,
$\nu \ne \bar {\nu }$), because an antineutrino is emitted in reaction
(1), whereas a neutrino is absorbed in reaction (4). I agree with
Pontecorvo [4] that Racah [3] did not consider neutrinoless
double beta decay (2$\beta (0\nu )$), but analyzed the
possibility of distinguishing Majorana and Dirac neutrinos in the
processes of inverse $\beta $ decay using free neutrino fluxes
(virtual intermediate state of a nucleus and virtual neutrino
were not considered!).\footnote {It is sometimes erroneously stated
that Racah [3] was the first who proposed the idea of 2$\beta (0\nu)$
decay.} In 1938, Furry [5] analyzed the ideas by Majorana
and Racah and arrived at rather pessimistic conclusions
concerning experimental possibilities of distinguishing Majorana
and Dirac neutrinos. These conclusions were primarily based on
the absence of intense sources of neutrinos at that time
(reactors were absent!). However, as early as in 1939, Furry [6]
considered for the first time the 2$\beta (0\nu)$ decay, i.e.,
the transformation of the ($A, Z$) nucleus to the ($A, Z + 2$)
nucleus accompanied by the emission of only two electrons. Furry
introduced the following scheme for describing the 2$\beta (0\nu )$
decay. The process occurs in two stages: (i) the initial
($A, Z$) nucleus emits one electron and transits to a virtual
intermediate state plus virtual $\bar {\nu }$ and (ii) this
virtual $\bar {\nu }$ as $\nu $ (since $\nu \equiv \bar {\nu }$)
is absorbed by the intermediate nucleus and induces its decay
with the emission of the second electron. This process can be
represented as follows:
\begin{gather}
\label{eq5}
(A, Z) \to (A, Z + 2) + 2e^{-}
\end{gather}

Thus, during five years (1935--1939), the main 2$\beta$ processes
were proposed and the first estimates of the lifetimes of nuclei
with respect to both 2$\nu $- (${\sim }10^{21}$--10$^{22}$ yr)
and 0$\nu $ decay (${\sim }10^{15}$--10$^{16}$ yr) were done. It was shown
that investigation of these processes can provide valuable
information on the properties of the neutrino. The main
contribution was made by M. Goeppert-Mayer and W. Furry.

\section{FIRST EXPERIMENTS (THE 1940s--1950s)}

At that time, the main motivation for experiments on the search
for 2$\beta$ decay was the possibility of determining the nature
of the neutrino (who is right, Dirac or Majorana?). The
theoretical estimates made at that time gave strongly different
values for these two possibilities: (0$\nu $ decay) ${\sim }10^{15}$
yr for the Majorana neutrino, whereas (2$\nu $ decay)
${\sim }10^{21}$ yr for the Dirac neutrino. It was clear that if
the neutrino has the Majorana nature, the 2$\beta(0\nu )$ decay
could be detected by detectors existing at that time.

In view of the World War II, active scientific investigations
resumed only in the second half of the 1940s. The first
experiment on the search for the 2$\beta $ decay was conducted in
1948 by Fireman [7] who sought the 2$\beta $ decay of $^{124}$Sn
using Geiger counters and obtained the limit $T_{1 / 2} > 3\times 10^{15}$~yr.
In 1949, Fireman carried out a new experiment with
$^{124}$Sn and obtained a positive result $T_{1 / 2} = (4{-}9)
\times 10^{15}$~yr [8]. Already in that pioneering experiment, he
used an enriched tin sample (25~g, enriched to 54${\%}$). This
result was not confirmed later in more sensitive experiments
performed in 1951--1953 [9--13]. The best bound for $^{124}$Sn
was ${T_{1 / 2} > 2 \times 10^{17}}$~yr [10]. In those years,
"positive" results often appeared, but they disproved by later
experiments. In particular, the "positive" effect was also observed
in $^{100}$Mo [14], $^{48}$Ca [15], and $^{96}$Zr [13]. In those
years, other 2$\beta $ decay processes with the emission of two
positrons (2$\beta ^{ + }$) and electron capture with the
simultaneous emission of a positron (EC$\beta ^{ + }$) [14, 16]
were sought. Winter [17] also pointed for the first time to the
possibility of a large increase in the probability of the
neutrinoless two-electron capture (ECEC(0$\nu $)) to an excited
level of the daughter nucleus under the resonance conditions
(zero transition energy). To search for the 2$\beta $ processes, the
most advanced methods and detectors were used. In particular,
Geiger, proportional, and scintillation counters, Wilson chamber,
and nuclear emulsion were used. Enriched isotopes ($^{48}$Ca,
$^{94}$Zr, $^{96}$Zr, $^{124}$Sn) were widely used. Instruments
were placed deep under the ground (to suppress the background
from cosmic rays); passive and active shieldings were used.
Nevertheless, the sensitivity of counter experiments at that time
was no more than ${\sim}10^{17}$--10$^{18}$ yr. Note that the
first experiment on the search for 2$\beta $ decay in the USSR
was conducted in 1956 [18].

In 1949, the first geochemical experiment was performed with
$^{130}$Te, which was transformed into $^{130}$Xe after 2$\beta $
decay [19]. This method consists of the separation of xenon from
ancient minerals (whose age is up to several billion years) and
their subsequent isotope analysis. The detection of an excess
amount of $^{130}$Xe (taking into account the contributions to
the effect from various nuclear reactions induced by neutrons,
cosmic rays, etc.) indicates the 2$\beta $ decay of the initial
nucleus and makes it possible to determine its half-life. The
sensitivity of the first geochemical experiment was much higher
than the sensitivities of counter experiments and a lower bound
of $8 \times 10^{19}$~yr was obtained. As early as in 1950,
Inghram and Reynolds [20] detected the 2$\beta $ decay of
$^{130}$Te and obtained $T_{1 / 2} = 1.4\times 10^{21}$~yr. This
result was initially not considered seriously, but it became
clear after 15--20 yr that 2$\beta (2\nu )$ decay was indeed
observed for the first time in this experiment. This occurred as
early as in 1950!

The first radiochemical experiment with $^{238}$U was also
carried out in 1950 [21]. The aim of that experiment was to seek
$^{238}$Pu ($T_{1 / 2} = 87.7$~yr), which should appear after the
2$\beta $ decay of $^{238}$U. To this end, plutonium was
separated from 14~kg of purified uranium oxide stored for six
years and the presence of $^{238}$Pu was detected by counting
$\alpha $ particles. As a result, the limit $T_{1 / 2} > 6
\times10^{18}$~yr was obtained.

The main achievement of that period (as was understood many years
later) was the first detection of 2$\beta (2\nu )$ decay in the
geochemical experiment with $^{130}$Te by Inghram and Reynolds.

\section{``MIDDLE AGES'' (THE 1960s--1970s)}

After the discovery of space parity violation in 1957 and the
determination of the $V$--$A$ nature of weak interaction (this
implies the presence of additional chiral selection), the
probability of observing 2$\beta (0\nu )$ decay decreased
strongly, because it became clear that the probability of such a
process can be much smaller than that for the 2$\beta (2\nu )$
transition. Nevertheless, interest in the 2$\beta $ processes
held in those years. In 1960, the probability of the 2$\beta
(0\nu )$ decay was calculated for the first time under the
assumption of the existence of the Majorana mass of the neutrino
[22]. Another possibility, namely, the admixture of right-handed
currents under the additional assumption of the identity of the
neutrino and antineutrino was simultaneously considered. At that
time, it became clear that the search for 2$\beta(0\nu )$ decay
is a good test for the lepton number conservation law.

Experiments at that time were scarce but of a very high level.
Mateosian and Goldhaber [23] achieved a sensitivity of $2 \times
10^{20}$ yr in the experiments with $^{48}$Ca; i.e., a
sensitivity of 10$^{20}$ yr was reached for the first time in
counter experiments. The working substance of the detector was
the enriched material under investigation (calcium fluoride
crystal containing 11.4~g of $^{48}$Ca enriched to 96.6${\%}$).
Thus, the "detector~$ \equiv$~studied substance" scheme was
implemented for the first time. In 1967, Fiorini et al. [24] for
the first time used a Ge(Li) detector to search for 2$\beta $ decay and
immediately obtained the bound $T_{1/ 2} > 3 \times 10^{20}$~yr
on the 2$\beta (0\nu )$ decay of $^{76}$Ge, which was the best
result at that time for counter experiments. Finally, the bound
${T_{1 / 2} > 5} \times 10^{21}$~yr fantastic for that time was
reached [25]. Correspondingly, wonderful prospects of the use of
germanium semiconductor detectors to search for 2$\beta $ decay were
demonstrated.

Almost at the same time, the research group led by Wu performed
two remarkable experiments at a setup consisting of a streamer
chamber in a magnetic field (visualization of tracks and
measurement of the energy) plus plastic scintillators
(simultaneity of the emission of electrons and the measurement of
the energy of electrons). At this setup, very stringent bounds on
2$\beta (0\nu )$ decay for isotopes with high energies of 2$\beta$
transitions were obtained: $T_{1 / 2} > 2 \times 10^{21}$~yr for $^{48}$Ca
(4.272~MeV)~[26, 27] and $T_{1 / 2} > 3.1 \times 10^{21}$~yr for
$^{82}$Se (2.995~MeV)~[28]. It is worth noting that the
sensitivity of counter experiments for the first time exceeded
10$^{21}$~yr in 1967 in the experiment with $^{48}$Ca [26].

It is worth noting that these results for $^{48}$Ca, $^{82}$Se,
and $^{76}$Ge in the early 1980s were considered as a lodestar.
All those who began to study 2$\beta $ decay at that time
compared their plans and dreams with those experiments.
Significant efforts were required in order to reach such a
sensitivity in new experiments. Even now (after 35--40~yr!), they
continue to be delightful.

In the second half of the 1960s, several research groups carried
out geochemical experiments with $^{130}$Te and $^{82}$Se:
research groups led by Takaoka [64] and Kirsten [118] confirmed the result
obtained in 1950 for $^{130}$Te and the research group led by
Kirsten observed the 2$\beta (2\nu)$ decay of
$^{82}$Se [29]. In 1975, the research group led by Manuel [30]
observed for the first time the 2$\beta $ decay of $^{128}$Te in
the geochemical experiment and determined the ratio $T_{1 /
2}$($^{130}$Te)$/T _{1 / 2}$($^{128}$Te).

The main achievements of that period are the sensitivity to
2$\beta (0\nu )$ decay above 10$^{21}$ yr in experiments with
$^{48}$Ca, $^{76}$Ge, and $^{82}$Se, the detection of 2$\beta
(2\nu)$ decay in geochemical experiments with $^{130}$Te,
$^{128}$Te, and $^{82}$Se. The main researchers were E. Fiorini,
C. Wu, T. Kirsten, and O. Manuel.

\section{"RENAISSANCE" (THE 1980s--1990s)}

Interest in 2$\beta$ decay was renewed in the early 1980s in view
of several circumstances. First, Lubimov et al. [31] stated the
detection of a neutrino mass of ${\sim}$30~eV in the experiment
performed at the Institute for Theoretical and Experimental
Physics to measure the $\beta $ spectrum of tritium (but this
statement was not confirmed in the subsequent experiments).
Second, the neutrino with a mass of several dozens eVs was
considered as a candidate for dark matter in cosmology. Third,
grand unified theories where the mass of the neutrino was
naturally treated as the Majorana mass were formulated. The appearance of the
Majorana mass means the breaking of the symmetry associated with
lepton charge conservation. This circumstance again enhanced
interest in the search for 2$\beta $ decay and initiated the
activity of theorists and experimenters.

In 1981, the following new type of 2$\beta $ decay, decay with
the emission of Majoron was considered [35]:
\begin{gather}
\label{eq6}
(A, Z) \to (A, Z + 2) + 2e^{-}+\chi ^{0}.
\end{gather}
The Majoron $\chi ^{0}$ is a massless Goldstone boson appearing
under the global breakdown of the $B$--$L$ symmetry, where $B$
and $L$ are the baryon and lepton quantum numbers, respectively.
If the Majoron exists, it can play a serious role in the history
of the early Universe and in the evolution of stars.

In 1982, the Schechter--Valle theorem was formulated [36]
according to which the observation of 2$\beta (0\nu )$ decay will
imply the existence of the Majorana mass of the neutrino in the
framework of gauge theories. This statement became a serious
theoretical reasons for experimenters to continue and improve
their experiments.

In 1985, the fundamental theoretical work by Doi, Kotani, and
Takasugi [37] appeared that remains an encyclopedia of the theory
of 2$\beta $ decay at present. In that work, the energy and
angular distributions of electrons were calculated with a good
accuracy for the mass mechanism and right-handed currents and the
possibility of distinguishing these mechanism in experiments
using the features of the corresponding distributions was
demonstrated.

The discrepancy between the theoretical predictions and
experiment bounds for 2$\beta (0\nu )$ decay was a serious
problem for a long time. Theories usually predicted the decay
rate one or two orders of magnitude higher than the existing
experimental bounds. However, in 1986, Vogel and Zirnbauer [38]
showed that the inclusion of the particle--particle interaction
in a nucleus allows a fairly accurate calculation of the rate of
2$\nu $ decay in the framework of the quasiparticle random phase
approximation (QRPA). For this reason, the QRPA models were
widely used to calculate nuclear matrix elements for both 2$\nu
$- and 0$\nu $ decays.

The activity of experimenters increased strongly. Several dozens
research groups began to search for 2$\beta $ decay. In the 1980s,
numerous measurements were performed with $^{76}$Ge using high
purity germanium (HPGe) detectors, primarily because large HPGe
detectors became available and quite cheap instruments. At the
same time, the choice of low-background construction materials,
the location of detectors surrounded by passive and active
shieldings deep under the ground provided the significant
reduction of the background (large contributions were made by
Avignone, Brodzinski, and Fiorini). All these led to an increase
in the sensitivity by several orders of magnitude. Several
research groups soon obtained the limits of 10$^{22}$ and
10$^{23}$~yr. The best experimental limit $T_{1 / 2} > 1.2 \times
10^{24}$~yr on the 2$\beta (0\nu )$ decay of $^{76}$Ge was
obtained by Caldwell [39] with the ${\sim }$7.2-kg high-purity
natural Ge detectors. In 1987, semiconductor Ge(Li) detectors
grown from enriched germanium were used for the first time
(ITEP--ErPhI experiment [40, 41]). All these achievements allowed
two large experiments with enriched germanium (Heidelberg--Moscow
[42] and IGEX [43]), in which a sensitivity of ${\sim
}10^{25}$~yr to 2$\beta (0\nu )$ decay was reached. A quiet
stringent limit on 2$\beta (0\nu )$ decay was also obtained for
$^{136}$Xe in the experiment with a time projection chamber with
3.3-kg xenon enriched in $^{136}$Xe to 62${\%}$. In addition to
the measurement of the total energy of electrons, events with the
simultaneous emission of two electrons from one points were
selected and tracks of electrons were reconstructed. 
After the measurements during
almost a year, the limit $T_{1 / 2} > 3.4\times 10^{23}$~yr was
obtained [44]. In 1984, Fiorini and Niinikoski [45] proposed
using low-temperature detectors to search for 2$\beta $ decay and the
Milano group successfully developed this method \footnote {It 
is necessary to notice that for the first time this idea has 
been stated in JINR (Dubna) by Mizelmaher, Neganov and Trofimov 
in 1982 [119],
but in JINR this idea was not realized.}.

The main experimental achievement in that period was the first
observation of the 2$\beta $(2$\nu $) decay of $^{82}$Se in the
direct counter experiment with a time projection chamber by the
research group led by Moe in 1987 [46]. Only 36 2$\beta $ events
were detected and the value $T_{1/2} = 1.1_{-0.3}^{+0.8} \times
10^{20}$~yr was obtained. This result was very important,
particularly from the psychological point of view. It removed a
certain invisible barrier. Positive results for 2$\nu$ decay
appeared as from a cornucopia. The 2$\beta (2\nu )$ decay of
$^{76}$Ge [41], $^{100}$Mo [47, 48], $^{116}$Cd [49--51], and
other nuclei was soon detected (at that time, 2$\beta(2\nu )$
decay was detected in direct experiments for seven nuclei). The
2$\beta (2\nu )$ decay of most nuclei was observed in several
independent experiments. The largest contribution was made by the
research group led by Moe ($^{82}$Se [46, 52], $^{100}$Mo [48],
$^{150}$Nd [53], and $^{48}$Ca [54]), by Ejiri et al. ($^{100}$Mo
[47] and $^{116}$Cd [49]), and in a series of the experiments
with the NEMO-2 detector ($^{100}$Mo [55], $^{116}$Cd [50, 56],
$^{82}$Se [57], and $^{96}$Zr [58]). Furthermore, the energy
spectra and angular distributions of electrons for all
investigated isotopes were studied in the NEMO-2 experiment. The
2$\beta (2\nu )$ decay of $^{238}$U was observed in a
radiochemical experiment in 1991 [59]. In the same year,
Barabash, Avignone, et al. detected for the first time 2$\beta
(2\nu )$ decay to the excited state of the
daughter nucleus ($^{100}$Mo--$^{100}$Ru($0_1^ + $; 1130.3~eV)
transition; the final result was reported in [60]).

Geochemical experiments were actively conducted in the 1980s and
almost stopped to the end of the 1990s. In the beginning of the
1980s, the half-lives of $^{130}$Te and $^{128}$Te were actively
discussed. Several research groups [61, 62] reported large
half-lives (${\sim }2.7 \times 10^{21}$ and ${\sim }7.7\times
10^{24}$~yr, respectively), whereas other authors [63--65]
obtained much smaller values (${\sim }0.8 \times 10^{21}$ and
${\sim }2 \times 10^{24}$~yr, respectively). Correspondingly, the
results differ by a factor of ${\sim }$3.5, although the stated
measurement accuracy was quite high (the accuracy of some
experiments reached 3${\%}$, as, e.g., in [62]). This problem was
not solved at that time. It was even assumed in [66] that this
discrepancy was probably caused by the time dependence of the
rate of 2$\beta $ decay (due, e.g., to time variations of the
weak interaction constant), because small $T_{1 / 2 }$ values
were obtained for "young" minerals (ages of ${\sim
}10^{7}$--10$^{8}$~yr), whereas large values were obtained for
"old" samples (ages of ${\sim }10^{9}$ yr or more). In 1993,
the first geochemical experiment with $^{96}$Zr was performed and
the half-life for the $^{96}$Zr--$^{96}$Mo transition was
obtained [67]. In this case, the daughter substance (molybdenum)
is not a gas in contrast to xenon and krypton, which are the
daughter elements in experiments with tellurium and selenium,
respectively.

The main achievements in that period are the first observation of
2$\nu$ decay in a direct (counter) experiment, the achievement of
sensitivity to 2$\beta (0\nu )$ decay in $^{76}$Ge higher than
10$^{25 }$~yr, the measurement of $T_{1 / 2}(2\nu )$ for ten
isotopes (in direct and geochemical experiments), and the first
observation of 2$\beta(2\nu )$ decay to the
excited state of the daughter nucleus. The most contribution was
made by the following researchers/collaborations: M.~Moe,
F.~Avignone, A.S.~Barabash, S.~Jullian, H.~Ejiri,
Yu.~G.~Zdesenko, H. Klapdor-Kleingrothaus, and
E.~Fiorini/Heidelberg--Moscow, IGEX, and NEMO-2. The contribution
of theorists S.~Petcov, O.~Civitarese, J.~Suhonen,
A.~Faessler,and F.~Simkovic (in addition to the researchers
listed above) should be mentioned.

\section{"CONTEMPORARY HISTORY" (FIRST DECADE OF THE 21st CENTURY)}

In 1998, the observation of neutrino oscillations in an
experiment with atmospheric neutrinos was reported at the
international conference "Neutrino-98" (Takayama, Japan). More
recently, neutrino oscillations were observed in experiments with
solar, reactor, and accelerator neutrinos. This means that the
neutrino has mass! In view of this circumstance, interest in
2$\beta (0\nu)$ decay in the beginning of the 21st century
increased strongly. It became clear that the detection and
investigation of 2$\beta (0\nu )$ decay can clarify many problems
of neutrino physics:

(i) The nature (Dirac or Majorana?) of the neutrino mass.

(ii) The absolute scale of the neutrino mass (measurement or
bound on the mass of the lightest neutrino).

(iii) Hierarchy type (normal, inverse, or quasidegenerate).

(iv) $CP$ violation in the lepton sector (measurement of the
Majorana $CP$-violating phases).

At the very beginning of that period, the Heidelberg--Moscow
Collaboration completed the measurements with $^{76}$Ge. The
interpretation of the experimental results was ambiguous. First,
the authors jointly reported the limit $T_{1 / 2} > 1.9 \times
10^{25}$~yr [42]. Then, Klapdor-Kleingrothaus et al. stated the
observation of 2$\beta (0\nu )$ decay of $^{76}$Ge with a
half-life of $1.5 \times 10^{25}$~yr [68]. The Moscow part of the
Collaboration disagreed with this statement [69]. Several years
later, Klapdor-Kleingrothaus changed the half-life to $1.19
\times 10^{25}$~yr [70], and after two years, he again changed it
to $2.23_{-0.31}^{+0.44} \times 10^{25}$~yr [71]. This
``observation'' was actively discussed and strongly criticized by
many physicists [72--75], and it is now clear that only new
experiments with $^{76}$Ge can clarify the situation.

Important results concerning the search for 2$\beta (0\nu )$
decay were obtained in the experiments CUORICINO [76,77,120]
(completed in 2008) and NEMO-3 [78--84] (completed in January 2011).
Those experiments involved large detectors with the masses of the
studied substances of 40 and 10~kg, respectively. The sensitivity
to 2$\beta (0\nu)$ decay reached in those experiments was higher
than 10$^{24}$~yr (>$2.8 \times 10^{24}$~yr for $^{130}$Te [120] and
>$1.1 \times 10^{24}$~yr for $^{100}$Mo [83]). In view of the
much larger phase-space volume of the studied nuclei ($^{130}$Te
and $^{100}$Mo), the sensitivity to the effective neutrino mass
was comparable with the sensitivity in experiments with
$^{76}$Ge. The NEMO-3 experiment simultaneously concerns seven
different isotopes and 2$\beta (2\nu )$ decay was detected for all these
isotopes. In addition to the precision measurement of $T_{1 /
2}(2\nu )$, this experiment includes the detection of all main
characteristics of 2$\nu $ decay: the total energy spectrum, the
energy spectrum of individual electrons and angular distribution.
To the end of 2008, more than 700\,000 (!) 2$\nu $ events were
detected for $^{100}$Mo against the almost zero background. The
$T_{1 / 2}(2\nu )$ value for $^{130}$Te was reliably measured for
the first time and, thereby, the old dispute between geochemists
was solved: the small current value of the half-life was
confirmed. In the same years, interest in 2$\beta ^{+ }$,
EC$\beta ^{ + }$, and ECEC processes increased and special
experiments were performed for the first time to seek the
resonance ECEC transition to an excited state of the daughter
nucleus (see review [85]).

Geochemical experiments were surprisingly revived at that time.
Experiments with $^{96}$Zr [86], $^{100}$Mo [87], $^{130}$Ba
[88], and $^{130}$Te [89, 90] were performed. As a result, the
ECEC(2$\nu $) process in $^{130}$Ba was observed for the first
time [88] and a serious attempt was made to explain the existing
discrepancies in the geochemical experiments with $^{130}$Te [89,
90]. At the same time, the results for $^{96}$Zr and $^{100}$Mo
are inconsistent with the results of the counter experiments.

The main achievements in that period are a controversial
situation with the "positive" result for $^{76}$Ge (can it be
called an achievement?), new results for 2$\beta (0\nu )$ decay
in $^{130}$Te and $^{100}$Mo, precision investigation of 2$\beta
(2\nu)$ decay for many nuclei, the first observation of the
ECEC(2$\nu$) process. The main contribution was made by H.
Klapdor-Kleingrothaus and Heidelberg--Moscow, CUORICINO, and
NEMO-3 Collaborations.

\section{CURRENT STATUS AND PROSPECTS}

Tables~1--3 present the best current data on 2$\beta (2\nu)$,
2$\beta (0\nu )$ and 2$\beta $(0$\nu \chi ^{0}$) decays. 
Tables~4--6 present the main historical landmarks
of the 75-yr investigations of these processes.

Thus, 2$\beta (2\nu )$ decay has been detected for ten nuclei
($^{48}$Ca, $^{76}$Ge, $^{82}$Se, $^{96}$Zr, $^{100}$Mo,
$^{116}$Cd, $^{128}$Te, $^{130}$Te,$^{150}$Nd, and $^{238}$U). In
addition, 2$\beta (2\nu )$ decay of $^{100}$Mo and $^{150}$Nd
to the $0_1^ + $ excited state of the daughter nuclei and
ECEC(2$\nu $) process in $^{130}$Ba were detected. Experiments on 2$\beta (2\nu
)$ decay achieved a qualitatively new level, where not only the
half-life, but also the other parameters of the process are
measured with a high accuracy (NEMO-3 experiment). At the same
time, the sensitivity of experiments on the search for 2$\beta $
decay with the transition to excited states of the daughter
nuclei, as well as 2$\beta ^{ + }$, EC$\beta ^{ + }$, and ECEC
processes, increases continuously. As a result, the transition to
thorough and comprehensive investigations of the 2$\nu $
processes is outlined, which provide very important information
on the nuclear matrix elements, the parameters of various
theoretical models, etc. A certain progress is observed in the
calculations of the nuclear matrix elements and in the
understanding of the nuclear physical aspects of 2$\beta $ decay,
although the accuracy of the calculations of the nuclear matrix
elements is still low.

Any 2$\beta (0\nu )$ decay has not been detected and, according
to Table~2, the conservative bound on the effective mass of the
Majorana neutrino is 0.7~eV. The current conservative bound on the
Majoron--neutrino coupling constant from the experiments on
2$\beta $ decay is $\langle g_{ee}\rangle < 1.7 \times 10^{-4}$.

Experiments of the next generation with the mass of the studied
isotopes ${\sim }$100--1000~kg will be launched in 
a few years and will provide a sensitivity of 0.1--0.01~eV to the
efficient mass of the Majorana neutrino , i.e., will allow the
analysis of the region of the predictions of the scheme with the
inverse neutrino-mass hierarchy. Table 7 presents the parameters
of several most promising planned experiments. First phase of GERDA 
(18 kg of $^{76}$Ge), EXO-200 (200 kg of $^{136}$Xe),
CUORE-0 ($\sim$ 40 kg of natural Te)
and KamLAND-Xe (400 kg of $^{136}$Xe) plan to start data-tacking in 2011.
For this reason one can expect occurrance of new, very interesting 
results in 2011-2012.

\section{CONCLUSIONS}

To conclude, it is worth noting that a significant contribution
to the study of the 2$\beta $ decay processes was made by
Soviet/Russian physicists. The first Soviet theoretical work was
published in 1950 by Sliv, who calculated the probability of
2$\beta $ decay [116]. In 1954, the first in world (of very high
quality!) review of the investigations of 2$\beta $ decay was
written by Zeldovich, Luk'yanov, and Smorodinsky [117]. From 1956
to 1965, Dobrokhotov, Lazarenko, and Luk'yanov performed several
high-quality (for that time) experiments with $^{48}$Ca (as was
mentioned in Section~3). In the 1980s, various Soviet research
groups conducted numerous experiments and obtained world-level
results (Baksan, Yerevan, Kiev, Moscow, Solotvino). At the end of
the 1980s--beginning of the 1990s, active cooperation with
foreign physicists began and joint experiments involving Russian
scientists were conducted in almost all best underground
laboratories of the world. One of the remarkable contributions of
the Russian researchers is the use of enriched isotopes produced
in the USSR/Russia. The largest experiments such as
Heidelberg--Moscow, IGEX, NEMO-2, and NEMO-3 experiments would be
impossible without Soviet/Russian scientists. At present, Russian
physicists participate almost in all large projects of future
experiments. Enriched isotopes (up to 1000~kg!) are planned to be
produced in Russia. In addition to those mentioned above, I list
only the leaders of Soviet/Russian research groups making
noticeable contribution to the investigation of 2$\beta $ decay:
A.~S.~Barabash, I.~V.~Kirpichnikov, and O.~Ya.~Zeldovich
(Institute for Theoretical and Experimental Physics, Moscow),
V.~B.~Brudanin (Joint Institute for Nuclear Research, Dubna,
Moscow region), V.~I.~Lebedev (Russian Research Centre Kurchatov
Institute, Moscow), Yu.~G.~Zdesenko (Institute for Nuclear
Research, National Academy of Sciences of Ukraine, Kiev), and
V.~V.~Kuzminov, A.~A.~Pomanskii, and A.~A.~Smolnikov (Institute
for Nuclear Research, Russian Academy of Sciences, Moscow).


\newpage
\begin{table}[h]
\setcaptionwidth{\linewidth} 
\setcaptionmargin{0mm}
\onelinecaptionsfalse 
\captionstyle{normal} %
\caption{Current $T_{1 / 2}$(2$\nu $) values (taken from [91])}
\begin{tabular}{l|r}
\hline
\multicolumn{1}{c|}{Isotope} &\multicolumn{1}{c}{$T_{1 / 2}$(2$\nu $), yr} \\
\hline
$^{48}$Ca &$4.4^{ + 0.6} _{-0.5 }\times 10^{19}$ \\
$^{76}$Ge&$(1.5\pm 0.1) \times 10^{21}$ \\
$^{82}$Se&$(0.92 \pm 0.07)\times 10^{20}$ \\
$^{96}$Zr&$(2.3\pm 0.2) \times 10^{19 }$ \\
$^{100}$Mo&$(7.1 \pm 0.4)\times 10^{18}$ \\
$^{100}$Mo--$^{100}$Ru(0$^{ + }_{1}$) &$5.9^{ + 0.8}_{ - 0.6} \times 10^{20 }$ \\
$^{116}$Cd&$(2.8 \pm 0.2)\times 10^{19}$ \\
$^{128}$Te&$(1.9 \pm 0.4)\times 10^{24}$ \\
$^{130}$Te&$6.8^{ + 1.2}_{-1.1}\times 10^{20}$ \\
$^{150}$Nd&$(8.2 \pm 0.9)\times 10^{18 }$ \\
$^{150}$Nd--$^{150}$Sm(0$^{ + }_{1}$)&$1.33^{ + 0.45}_{ - 0.26 }\times 10^{20}$ \\
$^{238}$U&$(2.0 \pm 0.6)\times 10^{21}$ \\
$^{130}$Ba, ECEC(2$\nu $)&$(2.2 \pm 0.5)\times 10^{21}$ \\
\hline
\end{tabular}
\end{table}

\begin{table*}[h]
\setcaptionwidth{\textwidth} 
\setcaptionmargin{0mm}
\onelinecaptionsfalse 
\captionstyle{normal} %
\caption{Best current results concerning the search for 2$\beta
$(0$\nu $) decay}
\begin{tabular}{l|l|l|c}
\hline \multicolumn{1}{c|}{Isotope} &
\multicolumn{1}{c|}{$E_{2\beta }$, keV}& \multicolumn{1}{c|}{$T_{1
/ 2}$ , yr}&
$\langle m_{\nu }\rangle $, eV \\
\hline
$^{48}$Ca&4272&>$5.8\times 10^{22}$~[99]&<14 \\
$^{76}$Ge&2039.0&>$1.9\times 10^{25}$~[42]&<0.22--0.66 \\
$^{82}$Se&2996&>$3.6\times 10^{23}$~[83]&<0.89--2.4\phantom{0} \\
$^{96}$Zr&3350&>$9.2\times 10^{21 }$~[84]&\phantom{0}<7.2--19.5 \\
$^{100}$Mo&3034.4&>$1.1\times 10^{24 }$~[83]&<0.45--0.93 \\
$^{116}$Cd&2805&>$1.7 \times 10^{23}$~[100]&<1.2--2.7 \\
$^{128}$Te&867&>$1.5\times 10^{24}$~(geochemistry)~[63, 91]&<1.7--4.3 \\
$^{130}$Te&2527.5&\phantom{0.}>$2.8\times 10^{24}$~[120]&<0.35--0.77 \\
$^{136}$Xe&2458.7&>$4.5\times 10^{23}$~[101]&<1.1--2.7 \\
$^{150}$Nd&3367&>$1.8\times 10^{22}$~[82]&<4.8--7.6 \\
\hline
\end{tabular}
\captionstyle{flushleft} \captionstyle{nonumber} \vspace{-5pt}
\caption{\footnotesize{Note. All bounds are given with 90${\%}$
C.L. The bounds on the effective mass of the Majorana neutrino
$\langle m_{\nu }\rangle $ were obtained using the calculated
nuclear matrix elements from [92--97] and phase-space volumes
from [98]. A decrease in the nuclear matrix elements for
$^{150}$Nd caused by the deformation of the nucleus is taken into
account (see [82]).\hfill}}
\addtocounter{table}{-1}
\end{table*}

\begin{table*}[h]
\setcaptionwidth{\textwidth} 
\setcaptionmargin{0mm}
\onelinecaptionsfalse 
\captionstyle{normal} %
\caption{Best results concerning the search for 2$\beta $(0$\nu \chi
^{0}$) decay ("odinary" Majoron)} 
\begin{tabular}{l|l|r}
\hline \multicolumn{1}{c|}{isotope} & \multicolumn{1}{c|}{$T_{1 /
2}$ , yr}&
\multicolumn{1}{c}{$\langle g_{ee}\rangle $} \\
\hline $^{48}$Ca& >$7.2\times 10^{20 }$~[102]& <$1.4\times 10^{ - 3}$ \\
$^{76}$Ge& >$6.4\times 10^{22 }$~[42]& <(0.79--2.3)~$\times ~10^{ - 4}$ \\
$^{82}$Se& >$1.5\times 10^{22 }$~[80]& <(0.64--1.7)~$\times ~10^{ - 4}$ \\
$^{96}$Zr & >$1.9\times 10^{21}$~[84]& <(1.5--5.7)~$\times ~10^{ - 4}$ \\
$^{100}$Mo& >$2.7\times 10^{22 }$~[80]& <(0.41--0.84)~$\times ~10^{ - 4}$ \\
$^{116}$Cd& \phantom{0.}>$8\times 10^{21 }$~[100]&
<(0.81--1.9)~$\times ~10^{ - 4}$ \\
$^{128}$Te& >$1.5\times 10^{24}$~(geochemistry)~[63, 91]&
<(0.67--1.7)~$\times ~10^{ - 4}$ \\
$^{130}$Te& >$2.2\times 10^{21 }$~[103]&
<(1.6--4.3)~$\times ~10^{ - 4}$ \\
$^{136}$Xe& >$1.6\times 10^{22}$~[101]&
<(0.87--2.4)~$\times ~10^{ - 4}$ \\
$^{150}$Nd& >$1.5\times 10^{21 }$~[82]&
<(1.7--3)~$\times ~10^{ - 4}$ \\
\hline
\end{tabular}
\captionstyle{flushleft} \captionstyle{nonumber} \vspace{-5pt}
\caption{\footnotesize{Note. All bounds are given with 90${\%}$
C.L. The bounds on the Majoron--neutrino coupling constant
$\langle g_{ee}\rangle $ were obtained using the calculated
nuclear matrix elements from [92--97] and phase-space volumes
from [98]. A decrease in the nuclear matrix elements for
$^{150}$Nd caused by the deformation of the nucleus is taken into
account (see [82]).\hfill}}
\addtocounter{table}{-1}
\end{table*}

\begin{table*}[h]
\caption{Main "landmarks" in double beta decay search.}
\begin{tabular}{|c|c|c|}
\hline
Date & Event & Remarks \\
\hline
1935 & Idea of 2$\beta2(\nu)$ decay   & M. Goeppert-Mayer [1] \\
1939 & Idea of 2$\beta0(\nu)$ decay   & W.H. Furry [6] \\
1948 & First experiment for search for 2$\beta$ & E.L. Fireman [7,8]; (Geiger counters and 25 g of \\
 & decay &   enriched $^{124}$Sn were used) \\
1950 & {\bf The first observation of }  & {\bf  M.G. Inghram, and J.H. Reynolds [20] } \\
&  {\bf 2$\beta2(\nu)$ decay} & {\bf (geochem. experiment with $^{130}$Te); } \\ 
& & {\bf $T_{1/2} \approx 1.4\cdot10^{21}$ y} \\
1966 & First counter experiment   & E. Mateosian, and M. Goldhaber [23]  \\
 & with sensitivity higher than $10^{20}$ y & ("detector=source", 11.4 g of enriched $^{48}$Ca);  \\
&  & $T_{1/2}(0\nu) > 2\cdot10^{20}$ y\\
1967 & First experiment with a Ge   & E. Fiorini et al. [24] (17 cm$^3$ Ge(Li) detector \\
 & semiconductor detector  & at the see level); $T_{1/2}(0\nu) > 3\cdot10^{20}$ y \\
1967 & Observation of 2$\beta(2\nu)$ decay  & T. Kirsten et al. [29] (geochemical experiment); \\
 & of $^{82}$Se & $T_{1/2} \approx 0.6\cdot10^{20}$ y \\
1967- & First counter experiment with  & R.K. Bardin, P.J. Gollon, J.D. Ullman, and \\
 1970 & sensitivity of higher than $10^{21}$ y & C.S. Wu [26,27] (strimmer chamber+    \\
 &  & scintillation counters); $T_{1/2}(0\nu$;$^{48}$Ca) $> 2\cdot10^{21}$ y,  \\
 &  & $T_{1/2}(2\nu$;$^{48}$Ca)$ > 3.6\cdot10^{19}$ y \\
1973 & High-sensitive counter experiment  & E. Fiorini et al. [25] (68 cm$^3$ Ge(Li) detector  \\
 & with $^{76}$Ge & at 4200 m w.e. depth); $T_{1/2}(0\nu$) $> 5\cdot10^{21}$ y \\
1975 & High-sensitive counter experiment   &	B.T. Cleveland et al. [28] (streamer chamber +  \\
 & with $^{82}$Se & scint. counters); $T_{1/2}(0\nu$; $^{82}$Se) $> 3.1\cdot10^{21}$ y \\
& & \\
1980- & Idea of 2$\beta$ decay with Majoron  & Models with singlet [32], doublet [33] and  \\
1981 & emission & triplet [34,35] Majoron were consided \\ 
1982 & Schechter-Valle theorem & J. Schechter, and J.W.F. Valle [36]; in gauge theories    \\
 &  & the detection of 2$\beta(0\nu)$ decay means that the neutrinos  \\
&  & has a mass and this mass is of the Majorana type \\

\hline

\end{tabular}
\end{table*}

\begin{table*}[h]
\caption{Main "landmarks" in double beta decay search (continuation of Table 4).}
\begin{tabular}{|c|c|c|}
\hline
Date & Event & Remarks \\
\hline
1984 & Program to develop low   & E. Fiorini, and T.O. Niinikoski [45]  \\
  & temperature detectors for double  & \\
& beta decay search & \\
1985 & Fundamental theoretical   & M. Doi, T. Kotani, and E. Takasugi [37] obtain  \\
  & analysis of 2$\beta$ decay  & the main  formulas for probability of decay, \\
 &  & energy and angular electron spectra  \\
1986 & $g_{pp}$ parameter of QRPA model  & P. Vogel, and M.R. Zirnbauer [38]; within frame- \\
 & (characterize the particle-particle   &  works of QRPA models the satisfactory \\
& interaction in nuclei)  & agreement between theoretical and experimental \\
&  & $T_{1/2}(2\nu)$ values for the first time has been observed \\
{\bf 1987} & {\bf First observation of 2$\nu$ decay  } & {\bf S.R. Elliott, A.A. Hahn, and M. Moe [46];}  \\
 & {\bf in a counter experiment} & {\bf TPC with $^{82}$Se; $T_{1/2}(2\nu)=1.1^{+0.8}_{-0.3}\cdot10^{20}$ y}\\
& &  \\
1987- & First counter experiment with a   & D.O. Caldwell et al. [39]; 8 detectors from natural   \\
1989 & sensitivity higher than $10^{24}$ y & Ge with a total weight of 7.2 kg;  \\
&   & $T_{1/2}(0\nu) > 1.2\cdot10^{24}$ y \\
1987- & First semiconductor detector  & ITEP-ErPI Collaboration [40,41]; 2 detectors \\
1990  & made of enriched germanium   & of enriched Ge with a total weight $\sim$ 1.1 kg).\\
 & (86\% of $^{76}$Ge)    & In 1990 $T_{1/2}(0\nu) > 1.3\cdot10^{24}$ y and \\
 &  &  $T_{1/2}(2\nu) = (0.9 \pm 0.1)\cdot10^{21}$ y were obtained [41]  \\
1991 & First observation of 2$\nu$ decay   & A.S. Barabash et al. [60]; low background HPGe \\
 & to the excited state of daughter & detector, 1 kg of $^{100}$Mo, $^{100}$Mo-$^{100}$Ru($0^+_1$;1130 keV)  \\
 & nucleus & transition; $T_{1/2}=6.1^{+1.8}_{-1.1}\cdot10^{20}$ y \\
1990- & Experiments with the   & H. Ejiri et al. [47,49]; observation of $2\beta(2\nu)$ decay  \\
 1998 & ELEGANT-V detector & in $^{100}$Mo and $^{116}$Cd \\
1991- & Experiments with the NEMO-2 & NEMO-2 Collaboration [55-58]; systematic   \\
1997 & detector  & investigation of $2\beta(2\nu)$ decay ($^{100}$Mo, $^{116}$Cd,   \\
 &  &  $^{82}$Se and  $^{96}$Zr) with the detection of  \\
& & all parameters of the decay \\

\hline
\end{tabular}
\end{table*}

\begin{table*}[h]
\caption{Main "landmarks" in double beta decay search (continuation of Table 5).}
\begin{tabular}{|c|c|c|}
\hline
Date & Event & Remarks \\
\hline

1991- & IGEX experiment & Measurements with 6.5 kg of enriched $^{76}$Ge;  \\
1999 &  & $T_{1/2}(0\nu) > 1.57\cdot10^{25}$ y [43] \\
1990- & Heidelberg-Moscow  & Measurements with 11 kg of enriched $^{76}$Ge [42]; \\
2003 & experiment & $T_{1/2}(0\nu) > 1.9\cdot10^{25}$ y,  \\
 &  & $T_{1/2}(2\nu)=1.74 \pm 0.01(stat)^{+0.18}_{-0.16}(syst)\cdot10^{21}$ y \\
2001 & First observation of ECEC(2$\nu$) & Geochemical experiment with $^{130}$Ba, \\
  &  & $T_{1/2} = (2.2 \pm 0.5)\cdot10^{21}$ y [88]\\
2002- & NEMO-3 experiment & NEMO-3 Collaboration [78-84];  \\
2010 &  & $T_{1/2}(0\nu$;$^{100}$Mo)$ > 1.1\cdot10^{24}$ y; observation and \\
 &  & precise investigation of $2\beta(2\nu)$ decay for 7 isotopes\\
 &  &   ($^{48}$Ca, $^{82}$Se, $^{96}$Zr, $^{100}$Mo, $^{116}$Cd, $^{130}$Te, $^{150}$Nd) \\
2003- & CUORICINO experiment & CUORICINO Collaboration [76,77,120]; \\
2008 &  & $T_{1/2}(0\nu$;$^{130}$Te)$ > 2.8\cdot10^{24}$ y \\

\hline
\end{tabular}
\end{table*}

\begin{table*}{}
\caption{Seven most developed and promising projects. 
Sensitivity at 90\% C.L. for three (1-st steps of GERDA and MAJORANA, KamLAND, SNO+), 
five (EXO, SuperNEMO and CUORE), and ten (full-scale GERDA and MAJORANA) 
years of measurements is presented. 
$^{*)}$ For the background 
0.001 keV$^{-1}\cdot kg^{-1} \cdot y^{-1}$; $^{**)}$ for the background 
0.01 keV$^{-1}\cdot kg^{-1} \cdot y^{-1}$. }
\begin{tabular}{|c|c|c|c|c|c|c|}
\hline
Experiment & Isotope & Mass of & Sensitivity & Sensitivity & Status & Start of  \\
& & isotope, kg & $T_{1/2}$, y & $\langle m_{\nu} \rangle$, meV & & data-tacking \\
\hline
CUORE [104,105] & $^{130}$Te & 200 & $6.5\cdot10^{26}$$^{*)}$ & 20-50 & in progress & $\sim$ 2013 \\ 
 & & & $2.1\cdot10^{26}$$^{**)}$ & 40-90 & & \\
GERDA [106,107] & $^{76}$Ge & 40 & $2\cdot10^{26}$ & 70-200 & in progress & $\sim$ 2012 \\
 & & 1000 & $6\cdot10^{27}$ & 10-40 & R\&D & $\sim$ 2015\\ 
MAJORANA [108,109] & $^{76}$Ge & 30-60 & $(1-2)\cdot10^{26}$ & 70-200 & in progress & $\sim$ 2013 \\
 & & 1000 & $6\cdot10^{27}$ & 10-40 & R\&D & $\sim$ 2015\\ 
EXO [110,111]  & $^{136}$Xe & 200 & $6.4\cdot10^{25}$ & 100-200 & in progress & $\sim$ 2011 \\
& & 1000 & $8\cdot10^{26}$ & 30-60 & R\&D & $\sim$ 2015\\ 
SuperNEMO [112-114] & $^{82}$Se & 100-200 & $(1-2)\cdot10^{26}$ & 40-100 & R\&D & $\sim$ 2013-2015\\
 & & & & & &\\
KamLAND [115] & $^{136}$Xe & 400 & $4\cdot10^{26}$ & 40-80 & in progress & $\sim$ 2011 \\
 & & 1000 & $10^{27}$ & 25-50 & R\&D & $\sim$ 2013-2015\\
SNO+ [121] & $^{150}$Nd & 56 & $4.5\cdot10^{24}$ & 100-300 & in progress & $\sim$ 2012 \\
 & &  500 & $3\cdot10^{25}$ & 40-120 &        R\&D  & $\sim$ 2015\\

\hline
\end{tabular}
\end{table*}


\begin{thebibliography}{99}
\addtolength{\itemsep}{3pt}
\bibitem{ 1} M.~Goeppert-Mayer, Phys. Rev. \textbf{48}, 512 (1935).
\bibitem{ 2} E.~Majorana, Nuovo Cimento \textbf{14}, 171 (1937).
\bibitem{3 } G.~Racah, Nuovo Cimento \textbf{14}, 322 (1937).
\bibitem{4 } B.~M.~Pontekorvo, Priroda, No. 1, 43 (1983).
\bibitem{ 5} W.~H.~Furry, Phys. Rev. \textbf{54}, 56 (1938).
\bibitem{ 6} W.~H.~Furry, Phys. Rev. \textbf{56}, 1184 (1939).
\bibitem{7 } E.~L.~Fireman, Phys. Rev. \textbf{74}, 1238 (1948).
\bibitem{ 8} E.~L.~Fireman, Phys. Rev. \textbf{75}, 323 (1949).
\bibitem{9 } J.~S.~Lowson, Phys. Rev. \textbf{81}, 299 (1951).
\bibitem{ 10} M.~I.~Kalkstein, and W.~F.~S.~Libby, Phys. Rev. \textbf{85}, 368 (1952).
\bibitem{ 11} R.~W.~Pearce and E.~R.~Darby, Phys. Rev. \textbf{86}, 1049 (1952).
\bibitem{ 12} E.~L.~Fireman, Phys. Rev. \textbf{86}, 451 (1952).
\bibitem{ 13} J.~A.~McCarthy, Phys. Rev. \textbf{90}, 853 (1953).
\bibitem{ 14} J.~H.~Fremlin and M.~C.~Walters, Proc. Phys. Soc. London, Sect. A \textbf{65}, 911 (1952).
\bibitem{15 } J.~A.~McCarthy, Phys. Rev. \textbf{97}, 1234 (1955).
\bibitem{ 16} A.~Berthelot et al., Compt. Rend. \textbf{236}, 1769 (1953).
\bibitem{17 } R.~G.~Winter, Phys. Rev. \textbf{99}, 88 (1955).
\bibitem{ 18} E.~N.~Dobrokhotov, V.~R.~Lazarenko, and S.~Yu.~Luk'yanov, Dokl. Akad. Nauk SSSR \textbf{110}, 966 (1956)
[Sov. Phys. Dokl. \textbf{1}, 600 (1956)].
\bibitem{19 } M.~G.~Inghram, and J.~H.~Reynolds, Phys. Rev. \textbf{76}, 1265 (1949).
\bibitem{ 20} M.~G.~Inghram, and J.~H.~Reynolds, Phys. Rev. \textbf{78}, 822 (1950).
\bibitem{ 21} C.~B.~Levine, A.~Griorso, and G.~T.~Seaborg, Phys. Rev. \textbf{77}, 296 (1950).
\bibitem{ 22} E.~Greuling and R.~C.~Whitten, Ann. Phys (N.Y.). \textbf{11}, 510 (1960).
\bibitem{23 } E.~der Mateosian and M.~Goldhaber, Phys. Rev. \textbf{146}, 810 (1966).
\bibitem{ 24} E.~Fiorini et al., Phys. Lett. B \textbf{25}, 602 (1967).
\bibitem{ 25} E.~Fiorini et al., Nuovo Cimento A \textbf{13}, 747 (1973).
\bibitem{ 26} R.~K.~Bardin, P.~J.~Gollon, J.~D.~Ullman, and C.~S.~Wu, Phys. Lett. B \textbf{26}, 112 (1967).
\bibitem{ 27} R.~K.~Bardin, P.~J.~Gollon, J.~D.~Ullman, and C.~S.~Wu, Nucl. Phys. A \textbf{158}, 337 (1970).
\bibitem{ 28} B.~T.~Cleveland, W.~R. Leo, C.~S. Wu, et al., Phys. Rev. Lett. \textbf{35}, 757 (1975).
\bibitem{ 29} T.~Kirsten, W.~Gentner, and O.~A.~Schaeffer, Z.~Phys. \textbf{202}, 273 (1967).
\bibitem{ 30} E.~W.~Hennecke, O.~K.~Manuel, and D.~D.~Sabu, Phys. Rev. C \textbf{11}, 1378 (1975).
\bibitem{ 31} V.~A.~Lubimov et al., Phys. Lett. B \textbf{94}, 266 (1980).
\bibitem{ 32} Y.~Chikashige, R.~N.~Mohapatra, and R.~D.~Peccei, Phys. Lett. B \textbf{98}, 265 (1981).
\bibitem{ 33} C.~S.~Aulakh and R.~N.~Mohapatra, Phys. Lett. B \textbf{119}, 136 (1982).
\bibitem{34} G.~B.~Gelmini and M.~Roncadelli, Phys. Lett. B \textbf{99}, 411 (1981).
\bibitem{ 35} H.~M.~Georgi, S.~L.~Glashow, and S.~Nussinov, Nucl. Phys. B \textbf{193}, 297 (1981).
\bibitem{ 36} J.~Schechter and J.~W.~F.~Valle, Phys. Rev. D \textbf{25}, 2951 (1982).
\bibitem{ 37} M.~Doi, T.~Kotani, and E.~Takasugi, Prog. Theor. Phys. Suppl. \textbf{83}, 1 (1985).
\bibitem{ 38} P.~Vogel and M.~R.~Zirnbauer, Phys. Rev. Lett. \textbf{57}, 3148 (1986).
\bibitem{39 } D.~O.~Caldwell, J.~Phys. G \textbf{17}, S137 (1991).
\bibitem{ 40} A.~A.~Vasenko et al., in \textit{Proceedings of the 2nd International Symposium
on Underground Physics'87, Baksan Valley, USSR, Aug. 17--19, 1987} (Nauka, Moscow, 1988), p. 288.
\bibitem{ 41} A.~A.~Vasenko et al., Mod. Phys. Lett. A \textbf{5}, 1299 (1990).
\bibitem{ 42} H.~V.~Klapdor-Kleingrothaus et al., Eur. Phys. J.~A \textbf{12}, 147 (2001).
\bibitem{ 43} C.~E.~Aalseth et al., Phys. Rev. C \textbf{65}, 09007 (2002).
\bibitem{ 44} R.~Luescher et al., Phys. Lett. B \textbf{434}, 407 (1998).
\bibitem{45 } E.~Fiorini and T.~O.~Niinikoski, Nucl. Instrum. Methods Phys. Res. \textbf{224}, 83 (1984).
\bibitem{ 46} S.~R.~Elliott, A.~A.~Hahn, and M.~K.~Moe, Phys. Rev. Lett. \textbf{59}, 2020 (1987).
\bibitem{ 47} H.~Ejiri et al., Phys. Lett. B \textbf{258}, 17 (1991).
\bibitem{ 48} S.~R.~Elliott, M.~K.~Moe, M.~A.~Nelson, and M.~A.~Vient, J.~Phys. G \textbf{17}, S145 (1991).
\bibitem{ 49} H.~Ejiri et al., J.~Phys. Soc. Jpn \textbf{64}, 339 (1995).
\bibitem{ 50} R.~Arnold et al., JETP Lett. \textbf{61}, 170 (1995).
\bibitem{ 51} F.~A.~Danevich et al., Phys. Lett. B \textbf{344}, 72 (1995).
\bibitem{ 52} S.~R.~Elliott et al., Phys. Rev. C \textbf{46}, 1535 (1992).
\bibitem{ 53} A.~De Silva et al., Phys. Rev. C \textbf{56}, 2451 (1997).
\bibitem{ 54} A.~Balysh et al., Phys. Rev. Lett. \textbf{77}, 5186 (1996).
\bibitem{ 55} D.~Dassi\'{e} et al., Phys. Rev. D \textbf{51}, 2090 (1995).
\bibitem{ 56} R.~Arnold et al., Z.~Phys. C \textbf{72}, 239 (1996).
\bibitem{ 57} R.~Arnold et al., Nucl. Phys. A \textbf{636}, 209 (1998).
\bibitem{ 58} R.~Arnold et al., Nucl. Phys. A \textbf{658}, 299 (1999).
\bibitem{ 59} A.~L.~Turkevich, T.~E.~Economou, and G.~A.~Cowan, Phys. Rev. Lett. \textbf{67}, 3211 (1991)
\bibitem{ 60} A.~S.~Barabash et al., Phys. Lett. B \textbf{345}, 408 (1995).
\bibitem{ 61} T.~Kirsten et al., in \textit{Proceedings of the International Symposium
on Nuclear Beta Decay and Neutrino (Osaka'86)} (World Sci., Singapore, 1986), p. 81.
\bibitem{ 62} T.~Bernatowicz et al., Phys. Rev. C \textbf{47}, 806 (1993).
\bibitem{ 63} O.~K.~Manuel, in \textit{Proceedings of the International Symposium
on Nuclear Beta Decay and Neutrino (Osaka'86)} (World Sci., Singapore, 1986), p. 71.
\bibitem{ 64} N.~Takaoka, and K.~Ogata, Z.~Naturforsch. A \textbf{21}, 84 (1966).
\bibitem{ 65} N.~Takaoka, Y.~Motomura, and K.~Nagano, Phys. Rev. C \textbf{53}, 1557 (1996).
\bibitem{ 66} A.~S.~Barabash, JETP Lett., \textbf{68}, 1 (1998); 
Eur. Phys. J. A \textbf{8}, 137 (2000); Astrophys. Space Sci. \textbf{283}, 607 (2003).
\bibitem{ 67} A.~Kawashima, K.~Takahashi, and A.~Masuda, Phys. Rev. C \textbf{47}, 2452 (1993).
\bibitem{ 68} H.~V.~Klapdor-Kleingrothaus et al., Mod. Phys. Lett. A \textbf{16}, 2409 (2001).
\bibitem{ 69} A.~M.~Bakalyarov et al., Phys. Part. Nucl. Lett. \textbf{2}, 77 (2005); hep-ex/0309016.
\bibitem{ 70} H.~V.~Klapdor-Kleingrothaus, I.~V.~Krivosheina, A.~Dietz, and O. Chkvorets, Phys. Lett. B \textbf{586}, 198 (2004).
\bibitem{ 71} H.~V.~Klapdor-Kleingrothaus and I.~V.~Krivosheina, Mod. Phys. Lett. A \textbf{21}, 1547 (2006).
\bibitem{ 72} C.~E.~Aalseth et al., Mod. Phys. Lett. A\textbf{ 17}, 1475 (2002).
\bibitem{ 73} Yu. G.~Zdesenko, F.~A.~Danevich, and V.~I.~Treyak, Phys. Lett. B \textbf{546}, 206 (2002).
\bibitem{ 74} A.~Strumia and F.~Vissani, Nucl. Phys. B \textbf{726}, 294 (2005).
\bibitem{ 75} O.~Chkvorets, arXiv:0812.1206 [hep-ex].
\bibitem{ 76} C.~Arnaboldi et al., Phys. Rev. Lett. \textbf{95}, 142501 (2005).
\bibitem{ 77} C.~Arnaboldi et al., Phys. Rev. C \textbf{78}, 035502 (2008).
\bibitem{ 78} R.~Arnold et al., Nucl. Instrum. Methods. A \textbf{536}, 79 (2005).
\bibitem{ 79} R.~Arnold et al., Phys. Rev. Lett. \textbf{95}, 182302 (2005).
\bibitem{ 80} R.~Arnold et al., Nucl. Phys. A \textbf{765}, 483 (2006).
\bibitem{ 81} R.~Arnold et al., Nucl. Phys. A \textbf{781}, 209 (2007).
\bibitem{ 82} J.~Argyriades et al., Phys. Rev. C \textbf{80}, 032501(R) (2009).
\bibitem{ 83} A.~S.~Barabash and V.~B.~Brudanin, Phys. At. Nucl. \textbf{74}, 312 (2011).
\bibitem{ 84} J.~Argyriades et al., Nucl. Phys. A \textbf{847}, 168 (2010).
\bibitem{ 85} A.~S.~Barabash, Yad. Fiz. \textbf{73}, 166 (2010) [Phys. At. Nucl. \textbf{73}, 162 (2010)]; arXiv:0807.2948 [hep-ex].
\bibitem{ 86} M.~E.~Wieser and J.~R.~De Laeter, Phys. Rev. C \textbf{64}, 024308 (2001).
\bibitem{ 87} H.~Hidaka, C.~V.~Ly, and K.~Suzuki, Phys. Rev. C \textbf{70}, 025501 (2004).
\bibitem{ 88} A.~P.~Meshik, C.M. Hohenberg, O.V. Pravdivtseva, and Y.S. Kapusta, Phys. Rev. C \textbf{64}, 035205 (2001).
\bibitem{ 89} A.~P.~Meshik, C.M. Hohenberg, O.V. Pravdivtseva, et al., Nucl. Phys. A \textbf{809}, 275 (2008).
\bibitem{ 90} H.~V.~Tomas et al., Phys. Rev. C \textbf{78}, 054606 (2008).
\bibitem{ 91} A.~S.~Barabash, Phys. Rev. C \textbf{81}, 035501 (2010).
\bibitem{ 92} M.~Kortelainen and J.~Suhonen, Phys. Rev. C \textbf{75}, 051303(R) (2007).
\bibitem{ 93} M.~Kortelainen and J.~Suhonen, Phys. Rev. C \textbf{76}, 024315 (2007).
\bibitem{ 94} F.~Simkovic et al., Phys. Rev. C \textbf{77}, 045503 (2008).
\bibitem{ 95} E.~Caurier et al., Phys. Rev. Lett. \textbf{100}, 052503 (2008).
\bibitem{ 96} J.~Barea and F.~Iachello, Phys. Rev. C \textbf{79}, 044301 (2009).
\bibitem{ 97} K.~Chaturvedi et al., Phys. Rev. C \textbf{78}, 054302 (2008).
\bibitem{ 98} J.~Suhonen and O.~Civitarese, Phys. Rep. \textbf{300}, 123 (1998).
\bibitem{ 99} S.~Umehara et al., Phys. Rev. C \textbf{78}, 058501 (2008).
\bibitem{100} F.~A.~Danevich et al., Phys. Rev. C \textbf{68}, 035501 (2003).
\bibitem{ 101} R.~Bernabei et al., Phys. Lett. B \textbf{546}, 23 (2002).
\bibitem{ 102} A.~S.~Barabash, Phys. Lett. B \textbf{216}, 257 (1989).
\bibitem{ 103} C.~Arnaboldi et al., Phys. Lett. B \textbf{557}, 167 (2003).
\bibitem{ 104} C.~Arnaboldi et al., Nucl. Instrum. Methods Phys. Res. A \textbf{518}, 775 (2004).
\bibitem{ 105} I.~C.~Bandac, J.~Phys. Conf. Ser. \textbf{110}, 082001 (2008).
\bibitem{ 106} I.~Abt et al., hep-ex/0404039.
\bibitem{ 107} J.~Janicsko-Csathy, Nucl. Phys. B Proc. Suppl. \textbf{188}, 68 (2009).
\bibitem{ 108} R. Gaitskell et al., Majorana White Paper, nucl-ex/0311013.
\bibitem{ 109} V.~E.~Guiseppe, arXiv:0811.2446 [nucl-ex].
\bibitem{ 110} M.~Danilov et al., Phys. Lett. B \textbf{480}, 12 (2000).
\bibitem{ 111} R.~Gornea, J.~Phys. Conf. Ser. \textbf{179}, 012004 (2009).
\bibitem{ 112} A.~S.~Barabash, Czech. J.~Phys. \textbf{52}, 575 (2002).
\bibitem{ 113} A.~S.~Barabash, Yad. Fiz. \textbf{67}, 2008 (2004) [Phys. At. Nucl. \textbf{67}, 1984 (2004)].
\bibitem{ 114} E.~Chauveau, AIP Conf. Proc. \textbf{1180}, 26 (2009).
\bibitem{ 115} K.~Nakamura, \textit{Report at International Conference Neutrino'2010, Athens, Greece, 13--19 June, 2010.}
\bibitem{ 116} L.~A.~Sliv, Zh. Eksp. Teor. Fiz. \textbf{20}, 1035 (1950).
\bibitem{ 117} Ya.~B.~Zeldovich, S.~Yu.~Luk'yanov, and Ya.~A.~Smorodinsky, Usp. Fiz. Nauk \textbf{54}, 361 (1954).
\bibitem{ 118} T.~Kirsten, W.~Gentner, and O.~Muller, Z. Naturforsch. \textbf{22a}, 1783 (1967). 
\bibitem{ 119} G.~V.~Mizelmaher, B.~S.~Neganov, V.~N.~Trofimov, Communication of JINR P8-82-549, Dubna, 1982 (in Russian).
\bibitem{ 120} E.~Andreotti et al., Astropart. Phys. \textbf{34}, 822 (2011).
\bibitem{ 121} J.~Maneira (SNO+ Collabotion), Neutrino Oscillation Workshop (Conca Soecchiulla, 
Italy, September 4-11, 2010).
\end{thebibliography}
\end{document}